\documentclass[aps,prl,twocolumn,amsfonts,showpacs,superscriptaddress]{revtex4} 
\usepackage{bm}
\usepackage{epsfig,amsopn}
\usepackage{graphicx}
\usepackage{amsmath,amssymb}
\usepackage{natbib}
\usepackage{color}
\usepackage{latexsym}
\usepackage{epsf}
\usepackage{amsfonts}
\usepackage{amssymb}

\renewcommand{\section}[1]{{\par\it #1.---}}
\def\beq{\begin{eqnarray}}
\def\eeq{\end{eqnarray}}
\def\etal{et al.}

\begin{document}
\title{
Kondo signature in heat transfer via a local two-state system
}

\author{Keiji Saito}
\affiliation{Department of Physics, Keio University, Yokohama 223-8522, Japan} 

\author{Takeo Kato}
\affiliation{Institute for Solid State Physics, the University of Tokyo, Kashiwa, Chiba 277-8581, Japan}

\date{\today}

\begin{abstract}
We study the Kondo effect in heat transport via a local two-state system.
This system is described by the spin-boson Hamiltonian with Ohmic dissipation, which
can be mapped onto the Kondo model with anisotropic exchange coupling.
We calculate thermal conductance by the Monte Carlo method based on the exact formula. 
Thermal conductance has a scaling form $\kappa = (k_B^2  T_K/\hbar) f(\alpha,T/T_K )$, 
where $T_K$ and $\alpha$ indicate the Kondo temperature and dimensionless coupling strength, respectively.
Temperature dependence of conductance is classified by the Kondo temperature as
$\kappa\propto (T/T_K )^3$ for $T\ll T_K$ and $\kappa\propto (k_B T / \hbar\omega_c)^{2\alpha-1}$ 
for $T\gg T_K$.
Similarities to the Kondo signature in electric transport are discussed.

\end{abstract}

\maketitle 
Heat and electric transport are ubiquitous phenomena.
Both phenomena have several similarities as well as dissimilarities.
Fourier's law in heat transport corresponds to Ohm's law in 
electric transport, and these laws are commonly categorized as diffusive transport. We also note that 
heat transport shows unique anomaly 
in low dimensions~\cite{BLR00,LLP03,dhar08}.
Ballistic transport leads to the quantization of conductance in 
electric~\cite{datta} as well as heat transport~\cite{rk98}.
The conductance quantum was measured in mesoscopic electric conduction in 1988~\cite{ex-elec}, and 
much later, the version of heat transport was also measured~\cite{ex-heat}.
Recently, the concept of thermal diode has also been discussed~\cite{baowen},
and an experiment has been conducted for demonstrating this~\cite{chang06}.
Recent progress in transport studies strongly indicates that 
heat transport analogue exists
for many categories of electric transport.

In this paper, we consider heat transfer between phononic reservoirs 
via a local two-state system, and aim to 
clarify the signature of the Kondo effect in heat transport.
This setup is analogous to electric transport via quantum dots, which is 
a typical and simplest example of quantum transport through a zero-dimensional physical object.
In quantum-dot systems, the Kondo effect is an interesting and famous phenomenon 
induced by electron correlation~\cite{datta,hewson,KondoExp}. 
The Coulomb blockade for electron tunneling is overcome by the formation of 
the Kondo singlet between a localized electric spin and conduction electrons.
Because of this effect,
electric conductance is nontrivially enhanced and 
can eventually reach conductance quantum.

\begin{figure}
\includegraphics[width=7.5cm]{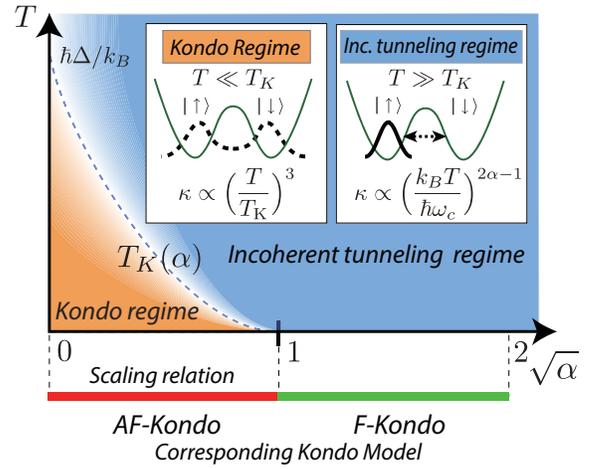}
\caption{(color online)
Schematic summary of results. 
The Kondo regime and the incoherent tunneling regime are shown 
in the space of temperature versus dimensionless coupling strength $\alpha$.
The dashed line indicates the Kondo temperature $T_K $, defined in Eq.~(\ref{kondo-temp}).
Mapping onto the corresponding parameter region of the Kondo model 
is shown below the figure; the region of $0\le \alpha <1$
corresponds to the Kondo model with antiferromagnetic exchange
coupling (AF-Kondo), and $1<\alpha\le 4$ to the one with ferromagnetic exchange
coupling (F-Kondo). 
The inset shows schematic quantum states
as well as main findings on thermal conductance (Eqs.~(\ref{lowtemp})-(\ref{kappa_highT})).
The scaling relation Eq.~(\ref{scaling}) holds for $0\le\alpha < 1$.}
\label{fig1}
\end{figure} 
We consider heat transfer via a two-level system modeled by the spin-boson system
with Ohmic dissipation~\cite{legget,weiss}.
In the equilibrium situation, low-energy physics of this model
is understood in terms of Kondo physics through mapping
onto the anisotropic Kondo model~\cite{legget,weiss,Guinea85}. 
Let $\alpha$ be a dimensionless 
coupling strength between the spin and bosonic reservoirs in the spin-boson Hamiltonian 
(exact definition is given below in Eq.~(\ref{spectrum1})).
The weak coupling region of $\alpha < 1$ is mapped onto
the antiferromagnetic parameter regime in the Kondo model, where 
the Kondo singlet is formed between a quantum dot and conduction electrons at sufficiently low temperatures.
On the other hand, the region of $\alpha > 1$ is mapped onto
the ferromagnetic parameter regime, in which only a trivial spin-doublet state is realized~\cite{suppl}.
At zero temperature, quantum states in the spin-boson model are clearly separated 
by quantum phase transition of the Kosterlitz-Thouless type 
at $\alpha=1$, where the local two-state forms one bonding state for $\alpha<1$ and
two degenerate states for $\alpha>1$. 
The Kondo temperature, a unique temperature scale characteristic of the Kondo effect, is
defined as a function of $\alpha$, and in the regime below the Kondo temperature 
(referred to as the Kondo regime), the local two-state is strongly and coherently correlated 
with the phononic environment. See Fig.~\ref{fig1} for schematic explanation.

The spin-boson model is a minimal model that describes molecular junctions \cite{nitzan}, 
a superconducting circuit~\cite{legget}, and a photonic waveguide with a local two-level 
system~\cite{LeHur12}, etc.; heat transport has also been studied intensively~\cite{vtw10,sn05,ro11,ns11}. 
These include thermal rectification effects~\cite{sn05}, 
the cotunneling process~\cite{ro11}, and fluctuations in current~\cite{ns11}. 
From the underlying Kondo physics in the equilibrium situation, it is also of general interest to analyze 
low-temperature properties in heat transport.
We note that the cotunneling mechanism has been studied~\cite{ro11}.
However, so far, no systematic studies have yet been conducted to get the
Kondo signature induced by higher-order processes beyond cotunneling.
This is the first study that shows the universal aspects arising from the underlying Kondo physics
based on the {\em exact} formula of heat current. We indicate similarities and dissimilarities of 
the Kondo signature between heat and electric transport via a zero-dimensional system.

\section{Model and exact current formula}
The local two-level system attached to
two phononic reservoirs is described by the following spin-Boson Hamiltonian:
\beq
H\!=\!{\hbar\Delta\over 2} \sigma_x +  \!\!\!
\sum_{\nu=L,R}\!\!\sum_{k} 
{\hbar\sigma_z\over 2}
\lambda_{\nu k} 
(b_{\nu k} + b_{\nu k}^{\dagger}) +\hbar \omega_{\nu k} b_{\nu k}^{\dagger} b_{\nu k}
\, ,\,   \label{hamiltonian1}
\eeq
where $\Delta$ is the tunneling frequency between the two states, i.e.,
the up-spin state $|\!\uparrow \,\rangle$ and the down-spin state $|\!\downarrow \,\rangle$.
The operator $\sigma_{\mu} \, (\mu=x,y,z)$ is the Pauli matrix, and
$b_{\nu k}$ is the annihilation operator of phonons with the wavenumber $k$
in the  $\nu$th reservoir.
The two reservoirs are characterized by the spectral density defined as
$I_{\nu } (\omega ) = \sum_{k} \lambda_{\nu k }^2 \, \delta (\omega - \omega_{\nu k} )$.
We assume the Ohmic dissipation for both reservoirs as follows: 
\beq
\begin{array}{ll}
I_{\nu} (\omega ) &= 2 \alpha_{\nu} \tilde{I} (\omega )  \, , 
~~~~~\alpha=\alpha_L + \alpha_R \, ,  \\
\tilde{I} (\omega ) &= \omega \,  \theta (\omega_c -\omega ) \, \theta (\omega ) \, ,
\end{array} \,  \label{spectrum1}
\eeq 
where $\alpha_\nu$ is the dimensionless coupling strength
between the system and the $\nu$th reservoir.
The cutoff energy $\hbar\omega_c$ is assumed to be sufficiently large compared to the
system's energy scale.
The Hamiltonian (\ref{hamiltonian1}) can be mapped onto the Kondo Hamiltonian
with anisotropic exchange coupling~\cite{Guinea85}. The in-plane and out-of-plane exchange 
parameters $J_{\parallel}$ and ${J_{\perp}}$ are related to the parameters in the spin-Boson
model as $\alpha = \left[ 1- {(2/\pi)}  \arctan (\pi \rho_0 J_{\parallel} / 4)  \right]^2 $ and 
$\Delta =\rho_0 \omega_c J_{\perp}$ respectively, where $\rho_0$ is a density of state 
(See the supplementary material \cite{suppl}).
The Kondo temperature $T_K$ is defined by renormalization group analysis~\cite{legget}:
\beq
T_K &=& \left\{  \begin{array}{ll}
{(\hbar / k_B)}  \, g_{\alpha}\Delta \left( {\Delta /  \omega_c }\right)^{\alpha/(1-\alpha)}  &\cdots~~ 
0\le \alpha < 1,  \\
0   &\cdots ~~  \alpha \ge  1,
\end{array}\right. ~~ \label{kondo-temp}
\eeq
where $k_B$ is the Boltzmann constant. The factor $g_{\alpha}$ is a nonuniversal constant, and 
throughout the paper, we take $g_{\alpha}=[\Gamma(1-2\alpha)\cos(\pi \alpha)]^{(1/2(1-\alpha))}$ with
the Gamma function $\Gamma(x)$~\cite{weiss}. 

The exact formula of heat current is derived with the standard procedure of the Keldysh formalism. 
The initial density matrix is prepared as the product form 
of equilibrium states of the system and left and right reservoirs. 
The temperatures of the left and right reservoirs are $T_L$ and $T_R$, respectively.
The heat current operator from the $\nu$th reservoir into the system is given by
\beq
J_{\nu} &=& i{\sigma_z \over 2} \sum_{k} \lambda_{\nu \, k} \hbar \omega_{\nu \, k} 
(- b_{\nu k}+ b_{\nu k}^{\dagger}) \, .
\eeq
Expressing this with the Keldysh green's function \cite{vtw10,wwz06}, and noting the 
Gaussian properties for bosonic variables, we derive a formal expression for 
current~\cite{saito}: $\langle J_{\nu} \rangle= {\hbar^2 /2} \int_{0}^{\infty} d \omega \omega
[ \chi'' (\omega ) I_{\nu}(\omega ) n_{\nu} (\omega ) -i C(\omega ) {I_{\nu} (\omega ) /2} ]$ where 
$C(\omega)$ is the Fourier transform of the lesser green function,
$C(t,t' )=-i \hbar^{-1} \langle \sigma_z (t')  \sigma_z (t) \rangle$, and 
$n_{\nu}(\omega)$ is the Bose-Einstein distribution of temperature $T_{\nu}$.
The symbol $\langle \cdots \rangle$ implies taking an average at a steady state.
The function $\chi'' (\omega )$ is the imaginary part of the Fourier's transform 
for the response function of the spin, which is defined as
$\chi(t,t') = i \hbar^{-1}\theta (t-t') \langle [\sigma_z (t) , \sigma_z (t') ]\rangle \, .$
Note that the well-known Landauer-type formula for the
ballistic transport cannot be derived because the spin-boson Hamiltonian is not bilinear.
However, we can derive an extended version of the Landauer-type formula
by considering zero dimensionality of the system and the conservation law 
of current $\langle J_L\rangle +\langle J_R\rangle =0$~\cite{mw92,ro08}. 
Then the exact heat-current formula is given by
\beq
\langle J_L\rangle \!\!
&=& \!\! {\hbar^2 \alpha \gamma  \over 4} \!
\int_{0}^{\infty} \!\!\! d \omega \,  \omega
\chi'' (\omega ) \tilde{I} (\omega) \left[ n_L(\omega )- n_R(\omega ) \right] , ~~~
\label{exact_formula}
\eeq
where $\gamma=4\alpha_L \alpha_R/\alpha^2$ is an asymmetry factor. 
In the linear response regime, thermal conductance defined by 
$\kappa = d\langle J_L\rangle / d T_L \,\vline_{\, T_L\to T_R=T}$ is given by
\beq
\kappa &=& {k_B \hbar \alpha \gamma \over 4}
\int_{0}^{\omega_c} \!\! d \omega \,  
S_{\alpha} (\omega )\, \omega^2
\Bigl[ {  \beta\hbar\omega/2  \over \sinh (\beta\hbar\omega /2 ) } \Bigr]^2 \!\!,~~~
\label{kappa}
\eeq
where we substitute the Ohmic spectral density for $\tilde{I}(\omega )$, and 
$S_{\alpha}(\omega)$ is the spectral function defined as
\beq
S_{\alpha}(\omega)&=&\chi''(\omega )/\omega \, .
\eeq
The formulas (\ref{exact_formula}) and (\ref{kappa}) are the basis of our calculation.
Before discussing the Kondo signature in heat transport, 
it is helpful to examine the current formula derived by the
quantum master equation approach by Segal and Nitzan~\cite{sn05}. 
By utilizing their expression of current, thermal conductance is obtained as
\beq
\kappa_{WC} &=& {k_B \alpha \gamma \over 4 }
 { \pi \Delta \over 2 n(\Delta ) +1 } 
\Bigl[ {  \beta\hbar\Delta /2  \over \sinh (\beta\hbar\Delta /2 ) } \Bigr]^2 \, ,~~~~
\label{wc}
\eeq
where $n(\omega)$ denotes the Bose-Einstein distribution of temperature $T$.
This expression is reproduced from the exact formula (\ref{kappa})
in the weak coupling limit ($\alpha \to 0$). This is checked by substituting 
the zeroth order of the spectral function, namely the expression for the isolated system:
$S_{0}(\omega) = \pi\delta(\omega-\Delta)/[\hbar\omega(2n(\omega ) +1)]$.

We note that at low temperatures $k_B T \ll \hbar \Delta$,
the weak coupling approximation predicts the Schottky-type temperature dependence
as $\kappa_{WC} \propto e^{-\hbar \Delta/k_B T}$, leading to the exponential suppression of heat current. 
This property is analogous to the Coulomb-blockade phenomenon in electric conduction, where
electric conductance is exponentially suppressed at low temperatures because of 
the excess electrostatic energy of electrons in a quantum dot.
However, as shown below, finite coupling to reservoirs remarkably changes 
the transport properties showing nontrivial universal properties.

\section{Numerical method}
\begin{figure}
\includegraphics[width=7.5cm]{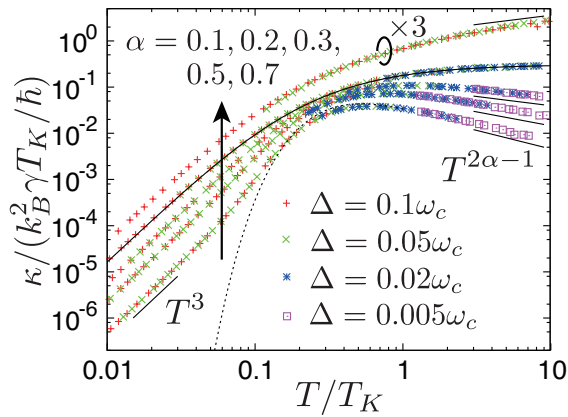}
\caption{(color online) Low-temperature behavior of the conductance
calculated with the Monte Carlo simulation. The data represent results for 
$\alpha=0.1,0.2,0.3,0.5$, and $0.7$ from bottom to top.
The solid line is an exact result for the Toulouse 
point $\alpha=1/2$. 
The figure clearly shows scaling form $\kappa (T) \sim (k_B^2 \gamma T_K/\hbar )\, f(\alpha,T/T_K)$
and the $T^3$-law in the low-temperature regime. The dotted line indicates 
$\kappa_{WC}$ for $\alpha=0.1$. For $T\gg T_K$, $\kappa$ depends on $T^{2\alpha -1}$. }
\label{fig2}
\end{figure}
In the subsequent sections, we focus on thermal conductance. 
In the exact formula (\ref{kappa}), the spectral function $S_{\alpha}(\omega)$
includes the entire information on the many-body effect.
We calculate $S_{\alpha}(\omega)$ by the Monte Carlo method, as follows. 
(i) We note that the path-integral representation of
the equilibrium partition function for the Hamiltonian (\ref{hamiltonian1}) is mapped 
onto a one-dimensional Ising model with long-range interaction
(for details, see the supplementary material \cite{suppl}).
(ii) For this Ising model, the Monte Carlo simulation with the Wolff algorithm~\cite{wolff}
is performed to obtain the spin-spin correlation, which is equivalent to Matsubara's green function 
${\cal G}(u)=\langle e^{uH} \sigma_z e^{-uH} \sigma_z \rangle_{\rm eq}$, where
$\langle \cdots \rangle_{\rm eq}$ implies an equilibrium average.
(iii) From the Fourier transformation of Matsubara's green function ${\cal G}(i \xi)$,
we calculate $\chi(\omega)$ using numerical analytic continuation 
$\chi(\omega)={\cal G}(i\xi\rightarrow\omega +  i\delta)$ 
by the Pad\'{e} approximation~\cite{volker,berezenski}.

\section{Kondo signature in thermal conductance}
In Fig.~\ref{fig2}, we show temperature dependence of the thermal conductance calculated 
with the abovementioned numerical procedure. 
The data indicate results for $\alpha <1$, which corresponds to
the antiferromagnetic coupling regime ($J_{\parallel} < 0$) in the mapped Kondo model.
The horizontal axis indicates the temperature scaled by the Kondo temperature 
$T_K$, and the vertical axis indicates the conductance scaled by $k_B^2 \gamma T_K/\hbar$.
To confirm numerical accuracy, we note that $\alpha=1/2$ 
is an exactly soluble point, called the Toulouse point, 
where the spectral function $S_{1/2}$ is given by~\cite{weiss}
\beq
S_{1/2} (\omega) &=& {\rm Im}\Bigl[ {4\over \hbar\pi \omega^2}{ k_B T_K \over \hbar\omega 
+ i (k_B T_K/\hbar )}  \left[ \psi (z' ) -\psi (z'') \right]\Bigr] \, ,
\nonumber 
\eeq
where $\psi(x)$ is the digamma function with the variables 
$z'=1/2 + \beta k_B T_K/(4\pi)$ and 
$z''=1/2 + \beta k_B T_K /(4\pi) -i \beta\hbar\omega/(2\pi)$. 
In Fig.~\ref{fig2}, we also show the exact values for the Toulouse point by a solid line. 
Our numerical results clearly agree with the exact results.
Another evidence of accuracy is given by a validation of the 
Shiba relation \cite{shiba75} for arbitrary coupling 
strength~\cite{suppl}.

Fig.~\ref{fig2} shows that all the data collapse onto
one curve for each value of $\alpha$ regardless of 
the tunneling frequency $\Delta$. This implies emergence of the scaling form in $\kappa$:
\beq
\kappa(T) &=& (k_B^2 \gamma T_K / \hbar )\, f(\alpha, T/T_K )\, . \label{scaling}
\eeq
This scaling form is an indication of the Kondo effect in heat transfer.
In addition, conductance is proportional to $T^3$ 
at sufficiently low temperatures $T \ll T_K$.
At high temperatures $T \gg T_K$, conductance depends on temperature as 
$T^{2\alpha-1}$, which intriguingly implies that 
the coupling strength determines monotonicity of temperature dependence.
These numerical findings are the first main results of this study.

For comparison, $\kappa_{WC}$ given by Eq.~(\ref{wc})
is shown for $\alpha=0.1$ in Fig.~\ref{fig2} by a dotted line. 
Although $\kappa_{WC}$ is quantitatively good around 
the Kondo temperature, it deviates from the
numerical results at lower temperatures, showing exponential reduction.
Enhanced heat transport from $\kappa_{WC}$ is analogous
to enhanced electronic transport via quantum dots in the Kondo regime.
However, note that conductance does not reach the universal quantum
of thermal conductance, $g(T)=\pi k_B^2 T/(6\hbar)$, which is linear in $T$. 
Heat transfer is generally sensitive to the scattering mechanism, and hence conductance 
tends to be reduced. This aspect is dissimilar to the Kondo signature in electric transport, 
where electric conductance can reach the conductance quantum.

\begin{figure}
\includegraphics[width=7.5cm]{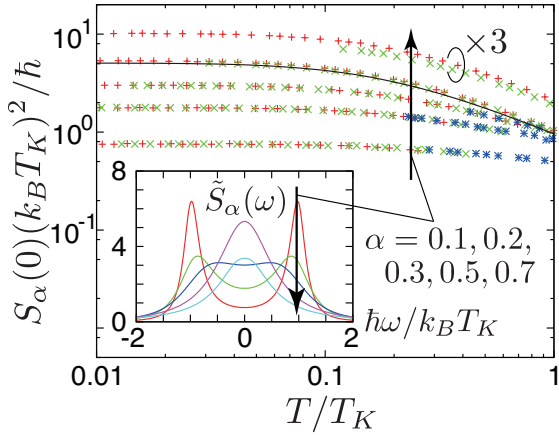}
\caption{(color online) Temperature dependence of $S_{\alpha}(0)$ for various $\alpha$
obtained by the Monte Carlo method. 
Legends are the same as those for Fig.~\ref{fig2}.
The solid line shows the analytic solution at the Toulouse point ($\alpha = 1/2$).
The inset shows $\omega$ dependence of the scaled spectral function
$\tilde{S}_{\alpha}(\omega)=S_{\alpha}(\omega)(k_BT_K)^2/\hbar$ for the low-temperature regime.}
\label{fig3}
\end{figure}  

To explain the universal $T^3$-law for $T\ll T_K$, we first consider the spectral function at the 
Toulouse point. The spectral function $S_{1/2}(\omega )$ is expanded with respect to temperature and frequency 
as in the form $S_{1/2} (\omega) =  \hbar / (k_B T_K)^2 [ 16/\pi 
+ O((\hbar\omega/ k_B T_K )^2) + O((T/T_K)^2 )]$, i.e., the second order
follows after the zeroth order.
In Fig.~\ref{fig3}, we show the temperature dependence of $S_{\alpha}(0)$. The inset 
shows $\omega$ dependence of $S_{\alpha}(\omega)$ for various $\alpha$ in the low-temperature regime. 
This figure shows that the spectral function generally has the same type of expansion 
as for the Toulouse point. 
Hence, the spectral function is safely approximated as 
$S_{\alpha} (\omega) \sim \hbar/(k_B T_K )^2 h_{\alpha}$
for a sufficiently low-temperature and small-frequency regime,
where $h_{\alpha}$ is a non-universal constant.
From this, conductance is approximated as
\beq
\kappa 
&\simeq&
{\hbar \alpha \gamma  
 \over 4 k_B T_K^2}
\int_{0}^{\infty} d \omega \,  
h_\alpha \, \omega^2
\Bigl[ {  \beta\hbar\omega/2  \over \sinh (\beta\hbar\omega /2 ) } \Bigr]^2 \nonumber \\
&\simeq & {(k_B^2 T_K {\cal N} / \hbar)} \, ( {T /  T_K} )^3  \,   ~~~{\rm for} ~ T \ll T_K \, ,
\label{lowtemp}
\eeq
where ${\cal N} =\pi^4 h_\alpha  \alpha\gamma/15$~\cite{comment2}.

We next consider the higher-temperature regime of $T_K  \ll  T \ll \hbar\omega_c/k_B$.
We start with the fluctuation dissipation theorem:
$S_{\alpha} (\omega )=D (\omega )/[\hbar\omega \coth(\beta\hbar\omega/2)]$,
where $D(\omega )$ is the Fourier transform of the symmetrized correlation function 
$D(t)= \langle \{ {\sigma}^z(t) ,{\sigma}^z \}\rangle_{\rm eq} /2-\langle \sigma^z\rangle_{\rm eq}^2$. 
Since the temperature is much higher than the Kondo temperature, the spin dynamics is incoherent and 
the correlation $D(t)$ decays exponentially. Hence, $S_{\alpha}(\omega)$ can be written as
$S_{\alpha} (\omega ) \simeq 2\zeta /[(\omega^2 + \zeta^2 )\hbar\omega\coth(\beta\hbar\omega/2)]$ 
where $\zeta$ is the decay rate of $D(t)$. 
We now employ the decay rate obtained using the noninteracting blip 
approximation~\cite{weiss85,fisher85,legget}; 
$\zeta \propto (\Delta^2/\omega_c)(\beta\hbar\omega_c)^{1-2\alpha}$.
By plugging this into the conductance formula (\ref{kappa}),
we get an approximation for conductance as 
\beq
\kappa &\simeq& {\cal C} \frac{k_B \Delta^2}{ \omega_c} 
\left(\frac{k_B T}{\hbar \omega_c}\right)^{2\alpha-1} \! \!
{\rm for} ~~T_K \ll T \ll \hbar\omega_c/k_B \, , ~~~~ \label{kappa_highT} 
\eeq
where ${\cal C}$ is a prefactor that weakly depends on coupling strength. 
This explains the numerical observation on temperature dependence 
in the high-temperature region $T_K \ll T$. 
Note that Eq.~(\ref{kappa_highT}) holds
at arbitrary temperatures for $\alpha > 1$ since the Kondo temperature is zero.

\begin{figure}
\includegraphics[width=7.5cm]{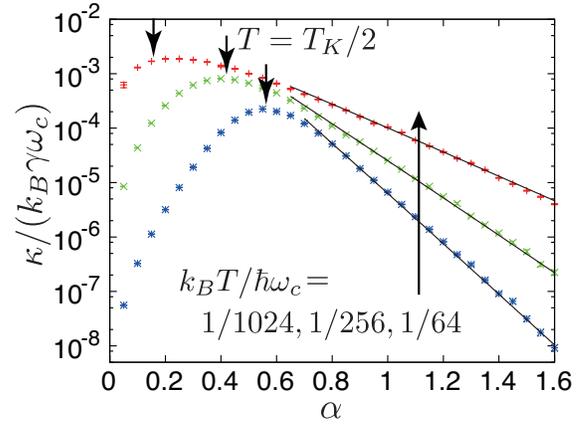}
\caption{(color online) Dependence of $\kappa$ on the couping strength $\alpha$ for three temperatures.
We set $\Delta = 0.05 \omega_c$. 
Conductance is numerically found to 
be maximum around $\alpha$ satisfying $T = T_K/2$, which is shown by small arrows.
The solid line shows the incoherent tunneling approximation~(\ref{kappa_highT}).
} 
\label{fig4}
\end{figure}  
We finally discuss coupling-strength dependence.
In Fig.~\ref{fig4}, conductance is shown as a function of $\alpha$ for fixed temperatures. 
As $\alpha$ increases, conductance first increases rapidly,
and then decreases exponentially. We note that the Kondo temperature is rapidly 
suppressed as $\alpha$ increases (see Eq.~(\ref{kondo-temp}) and Fig.~\ref{fig1}).
The former increase corresponds to the Kondo
regime ($T \ll T_K(\alpha)$), and is due to the suppression of $T_K$ (see Eq.~(\ref{lowtemp})).
On the other hand, the latter exponential decrease 
corresponds to the incoherent tunneling regime ($T \gg T_K(\alpha)$), and
is caused by the factor $(k_BT/\hbar\omega_c)^{2\alpha-1}$ in Eq.~(\ref{kappa_highT}).
The exponential reduction of conductance is an evidence of
the strong suppression of tunneling due to large dissipation,
consistently with an indication in \cite{vwt08,nikolin11}. 
Clear exponential suppression explained by Eq. (\ref{kappa_highT}) is the second main result.

\section{Summary}
We considered thermal transport via a local two-state system,
and investigated low-temperature properties on the basis of the exact formula~(\ref{kappa}).
Our findings are schematically summarized 
in Fig.~\ref{fig1}. We obtained, as Kondo-signature, the scaling form (\ref{scaling}),
and conductance classified by the Kondo temperature;
Eq.~(\ref{lowtemp}) for $T\ll T_K$ and Eq.~(\ref{kappa_highT})
for $T_K \ll T \ll \hbar\omega_c / k_B$.
The exponential reduction of conductance for large coupling 
is explained by the formula Eq.~(\ref{kappa_highT}). 
We hope that our study motivates further research on low-energy heat transfer via local systems.
Effect of bias, other types of dissipation~\cite{ks} and far-from-equilibrium effect, e.g., for problems in 
\cite{segal13,ren13}
will be intriguing subjects in this direction.

\noindent\\
{\bf Acknowledgement\hfill} \\
We are grateful to Mikio Eto, Teemu Ojanen, Rui Sakano and Dvira Segal
for useful comments. 
We also thank Ulrich Weiss for showing the derivation 
of $g_{\alpha}$. 
KS was supported by MEXT (23740289).
TK was supported by JSPS KAKENHI Grant Number 24540316.

\begin{widetext}
\begin{center}
Supplementary Material for \protect \\ 
``Kondo signature in heat transfer via a local two-state system''
\end{center}
\end{widetext}

\section{Anisotropic Kondo model}
The Hamiltonian of the anisotropic Kondo model is given by 
\beq
H_{AK} &=& \sum_{k}\sum_{\sigma=\uparrow , \downarrow} \epsilon_k c_{k \, \sigma}^{\dagger} c_{k\, \sigma} \nonumber \\
&+& {J_{\perp}} \sum_{k ,k'} ( c_{k \uparrow}^{\dagger} c_{k' \downarrow} S^{-}
+ c_{k \downarrow}^{\dagger} c_{k' \uparrow} S^{+}  )
\nonumber \\
&+& {J_{\parallel} \over 2}\sum_{k, k'} 
( c_{k \uparrow}^{\dagger} c_{k' \uparrow} -
c_{k \downarrow}^{\dagger} c_{k' \downarrow}
) S_z \, ,
\eeq
where $c_{k\, \sigma}$ is an annihilation operator of the fermion of the wave number $k$ and spin $\sigma$. Operators 
$S_{\mu} (\mu=x,y,z)$ is the spin operator by localized electron inside quantum-dot and 
$S^{\pm}=S_{x}\pm i S_{y}$. The parameters
$J_{\parallel}$ and ${J_{\perp}}$, respectively, represent the in-plane and out-of-plane exchange parameters.
When $J_{\parallel} > 0$, the Hamiltonian describes the antiferromagnetic Kondo (AF-Kondo) model, 
while $J_{\parallel} < 0$ corresponds to the ferromagnetic Kondo (F-Kondo) model. 
The correspondence between the anisotropic Kondo model and 
the spin-boson model via bosonization~\cite{guinea-s,costi-s} implies 
\beq
\alpha &=& \Bigl[ 1- {2\over \pi}  \arctan (\pi \rho_0 J_{\parallel} / 4)  \Bigr]^2 \, ,\\
\Delta &=&\rho_0 \omega_c J_{\perp} \, , 
\eeq 
where $\rho_0 $ is the density of the state of conduction electrons. 
These relations indicate that the regime $-\infty < \rho J_{\parallel} < \infty$ is mapped onto
$4 > \alpha > 0$. The phase transition point $\alpha =1$ in the spin-boson model
corresponds to the transition point between
the AF-Kondo and F-Kondo regime (see Fig.~1 in the main text).

\section{Exact formula of heat transfer}
The formal current formula is given by
\beq
\langle J_{\nu} \rangle
\!\!=\! {\hbar^2 \over 2 } 
\int_{0}^{\infty} \!\! d \omega \omega
[ \chi'' (\omega ) I_{\nu}(\omega ) n_{\nu} (\omega ) -
i C (\omega ) {I_{\nu} (\omega ) /2} ]\, .~~ 
\eeq
We define a quantity 
\beq
r_{\nu} &=& 
{\int_{0}^{\infty} d\omega \hbar \omega  C (\omega ) I_{\nu} (\omega ) 
\over 
\sum_{\nu'=L,R}
\int_{0}^{\infty} d\omega  \hbar \omega  C (\omega ) I_{\nu '} (\omega ) } \, .
\eeq
From the conservation law $\langle J_L\rangle+\langle J_R \rangle=0$, the current formula is rewritten as
\beq
\langle J_L\rangle &=& r_R \langle {\cal J}_L\rangle - r_L \langle {\cal J}_R\rangle 
 \nonumber \\
&=& {\hbar\over  2 }
\int_{0}^{\infty} d \omega \, \hbar \omega  \, 
 \chi'' (\omega ) \nonumber \\
&\times&
\Bigl[ r_R I_L (\omega ) n_L(\hbar\omega )-r_L I_R (\omega ) n_R(\hbar\omega ) \Bigr] \, . 
\label{mw:formula}
\eeq
Note that if $\tilde{I}_{L} (\omega)=\tilde{I}_{R}(\omega) =\tilde{I} (\omega )$, 
$r_{\nu}$ is simplified as $r_{\nu} = \alpha_{\nu}/( \alpha_L + \alpha_R )$. Hence, the above formula is 
reduced to (5) in the main text. 

\section{Weak coupling limit ($\alpha \rightarrow 0$)}
Let $H_s$ be the system's Hamiltonian, i.e., $H_{s}= {\hbar \Delta \sigma_x/ 2}$.
It is diagonalized as
\beq
e^{i \pi \sigma_y /4 }  H_{s} e^{-i \pi \sigma_y /4 } 
&=& {\hbar \Delta \sigma_z / 2 }  \, .
\eeq
We first express Segal and Nitzan's result \cite{sn05-s}, 
which is obtained from the quantum master equation:
\beq
\langle J \rangle =
{\hbar\pi \over 2 }{ \Delta I_{L}(\Delta ) I_{R} (\Delta )  \left[ n_L (\Delta ) - n_R (\Delta )  \right] 
\over 
I_L (\Delta ) (2 n_{L} (\Delta) + 1 ) + I_R (\Delta ) (2 n_{R} (\Delta) + 1 )   } \, . \label{ffe} ~~~
\eeq
Note that the above expression is the first-order expression with respect to 
the coupling strength between the system and reservoirs. Then, we find the zeroth-order expression for $\chi''$, 
which implies the response function at the weak coupling limit. For the case of thermal conductance,
we use the equilibrium value of $\chi ''$:
\beq
\chi '' (\omega ) &\to & 
{\pi / [\hbar ( 2 n(\omega ) +1)] } \, \delta (\omega - \Delta ) \, . \label{equilibrium}
\eeq
We note that contribution arises only from $\omega=\Delta$, which immediately implies
$r_{\nu} = I_{\nu} (\Delta )/(I_{L}(\Delta) + I_R(\Delta) )$. 

To obtain the expression at far-from-equilibrium (\ref{ffe}), we replace the temperature in Eq.(\ref{equilibrium})
with an effective temperature at the weak coupling limit between the reservoirs:
\beq
\chi '' (\omega ) &\to & 
{\pi / [\hbar ( 2 n_{\rm eff} (\omega ) +1)] }\, \delta (\omega - \Delta ) \, . 
\eeq
In order to obtain the effective temperature, we use the quantum master equation.
We write the master equation in the representation diagonalizing the spin Hamiltonian:
\beq
\dot{\rho} &=& i \left[ \rho , {\Delta \sigma_z / 2 } \right]
-\!\!\sum_{\nu=L,R} 
\left\{ 
\left[ X , R_{\nu} \rho \right] + \left[ X , R_{\nu} \rho \right]^{\dagger}
\right\} \, ,~~~~~ \\
X &=& - \sigma_x \, ,
\eeq
where the matrices $R$ are given by the $2\times 2$ matrix:
\beq
R_{\nu} &=& -
I_{\nu} (\Delta )
\left( 
\begin{array}{cc}
0 , & n_{\nu } (\Delta ) \\
- n_{\nu } ( - \Delta ) , & 0 
\end{array}
\right) \, .
\eeq
In case of a single reservoir, the equilibration is guaranteed by the detailed balance
for the matrix $R_{\nu}$:
\beq
R_{\nu \, 1,2} /R_{\nu \, 2,1} 
= n_{\nu} (\Delta ) / [- n_{\nu} (- \Delta )] = e^{-\beta_{\nu} \hbar \Delta} \, .
\eeq
Then, by analyzing the effective detailed balance for the nonequilibrium situation with two 
reservoirs, the effective temperature is obtained as
\beq
e^{-\beta_{\rm eff} \hbar \Delta } &=& 
{I_L (\Delta ) n_L (\Delta) + I_{R} (\Delta ) n_R (\Delta ) \over 
-I_L (\Delta ) n_L (-\Delta) - I_{R} (\Delta ) n_R (- \Delta ) } \, .~~~~~
\eeq
From this, a simple manipulation yields
\beq
\chi '' (\omega ) &=& 
{
I_L (\Delta )  + I_{R} (\Delta )
\over
I_L (\Delta ) [2 n_L (\Delta) + 1] + I_{R} (\Delta ) [ 2 n_R (\Delta ) +1 ]
} \nonumber \\
&\times&{\pi  / \hbar } \, \delta (\omega - \Delta ) \, .
\eeq
Combining the ratio 
$r_{\nu} = I_{\nu} (\Delta )/(I_{L}(\Delta) + I_R(\Delta) )$, 
one gets the result (\ref{ffe}).

\section{Ising model with long-range interaction}
The spin-boson model can be mapped onto an Ising model with long-range exchange
interaction. We start with the derivation of the partition function for the spin-boson model.
For simplicity, we set $\hbar=1$. We first divide the Hamiltonian into two parts and 
create the following definition:
\beq
H &=& \Delta\sigma_x /2 + H_z \, , \\
H_z &=& 
{\sigma_z\over 2}\sum_{\nu=L,R}\sum_{k} \lambda_{\nu k} 
(b_{\nu k} + b_{\nu k}^{\dagger}) 
\nonumber \\
&&+ \sum_{\nu=L,R}\sum_{k} \omega_{\nu k} 
b_{\nu k}^{\dagger} b_{\nu k} \, , ~~~~ \\ 
\tilde{\sigma}_{x} (u) &:=& e^{iH_z u } \sigma_x e^{-iH_z u } \, .
\eeq
Then, the partition function $Z = {\rm Tr} e^{-\beta H}$ is divided into two parts: 
$Z=Z_{+}+Z_{-}$ as
\beq
Z &=& Z_{+}+Z_{-} \nonumber \\
Z_{\pm} &=&  \langle \pm | 
{\rm Tr}_{\rm boson } e^{-\beta H} | \pm \rangle \, ,
\eeq
where $|\pm \rangle$ is the eigenstate of $\sigma_z$, i.e., 
$\sigma_z |\pm \rangle = \pm 1\, |\pm \rangle$, 
and ${\rm Tr}_{\rm boson}$ implies the trace with respect to boson's degrees of freedom.
We expand $Z_{+}$ as
\beq
Z_+ &=& {\rm Tr}_{\rm boson} \Bigl\{
  \langle + |  e^{-\beta H_z} e_{\leftarrow}^
{
-\int_{0}^{\beta} du {\Delta\tilde{\sigma_x}(u) / 2 }
}|+\rangle
\Bigr\}   \, \nonumber \\
&=&
\sum_{n=0}^{\infty} 
{\rm Tr}_{\rm boson} \Bigl\{
  \langle + |  
e^{-\beta H_z} 
\int_{0}^{\beta }\!\! d\tau_1 \cdots\int_{0}^{\tau_{2n-1}-\tau_c} \!\! d \tau_{2n }  \nonumber \\ 
&\times& \left( {\Delta \over 2} \right)^{2n} 
\tilde{\sigma}_x (\tau_1) \cdots\tilde{\sigma}_x (\tau_{2n})
|+\rangle
\Bigr\}   ,
\eeq
where $\tau_c =1/\omega_c$.
By taking trace with respect to the boson part, 
we obtain a formal expression for $Z_+$~\cite{leggett-s,volker-s}:
\beq
Z_{+} &=& Z_0
\sum_{n=0}^{\infty} \left( {\Delta \tau_c \over 2} \right)^{2n} 
\int_{0}^{\beta }\!\! {d\tau_1\over \tau_c} \cdots \int_{0}^{\tau_{2n-1}-\tau_c} \!\! {d \tau_{2n }
\over \tau_c}  \nonumber \\
&\times& 
\exp\Bigl\{ 
2 \alpha\sum_{i < j} (-1)^{i+j}
\ln \Bigl| {\beta \over \pi\tau_c } \sin (\pi(\tau_j - \tau_i)/\beta )  \Bigr|
\Bigr\} \, ,\nonumber \\
\eeq
where $Z_0$ is a partition function for the free boson part, and $Z_-$ takes the same form.
Finally, we connect this to the kink dynamics in the Ising model~\cite{volker-s,cardy-s}. 
The equivalent Ising dynamics is obtained
when the imaginary time is considered to be a position of spin.
By discretizing the space, the equivalent Ising Hamiltonian $H_I$ reads
\beq
\beta_I H_I &=& -{J_{nn}\over 2} \sum_{i=1}^{N} \sigma_i \sigma_{i+1} 
- {\alpha\over2} \sum_{j< i} {(\pi/N)^2 \sigma_i \sigma_j 
\over \sin^2 \left[\pi (j-i)/ N\right] } \, ,~~~~~~
\eeq
where $\beta_I$ is the inverse temperature in the mapped Ising model. 
The nearest neighbor interaction coefficient
$J_{nn}$ is given by $J_{nn}=-\alpha (1+\gamma ) -\ln (\Delta\tau_c /2)$,
where $\gamma$ is Euler's constant. The lattice number $N$ is determined by $N=\beta\omega_c$, 
and hence the Monte Carlo simulation is possible only for $\beta\omega_c \gg 1$. 
\begin{figure}
\includegraphics[width=8cm]{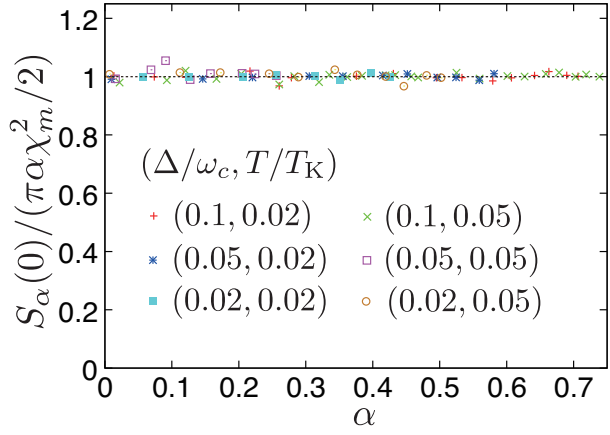}
\caption{(color online) Numerical validation of the Shiba relation.}
\label{shiba}
\end{figure}  

\section{The Shiba relation and the cotunneling formula}
The Shiba relation~\cite{shiba-s75}, which is a general identity equation valid at low temperatures
well below the Kondo temperature, is written by
\beq
S_{\alpha} (0) &=& {\alpha \pi \chi_m^2 \over 2} \, ,
\label{eq:ShibaRelation}
\eeq
where $\chi_m$ is the susceptibility of the local spin.
For the numerical validation, we calculate $S_{\alpha}(0)$ and $\chi_m$
at temperatures much lower than $T_K$, and show the ratio
$S_{\alpha}(0)/(\alpha \pi \chi_m^2/2)$ as a function of $\alpha$ 
for several sets of $(\Delta,T)$ in Fig.~\ref{shiba}.
The figure clearly demonstrates the validity of the Shiba relation, implying
one of the evidences on the reliability of our numerical calculations.

By substituting the Shiba relation (\ref{eq:ShibaRelation}) into 
the thermal conductance (Eq.~(6) in the main text), 
we obtain an approximation which is valid at low temperatures ($T \ll T_K$)
\beq
\kappa \sim \frac{\pi k_B \hbar^2 \chi_m^2}{8}  \int_0^{\omega_c} d\omega
I_L(\omega) I_R(\omega) \left[\frac{\beta \omega/2}{\sinh(\beta \omega/2)} \right]^2 \, .
\eeq
For weak coupling ($\alpha_L, \alpha_R \ll 1$), the susceptibility $\chi_m$
is approximately given by $2/\hbar \Delta$, which is the one for the isolated 
two-state system, and one can recover the cotunneling formula derived in Ref.~\cite{Ruokola11}. 
For arbitrary coupling, however, the cotunneling formula of Ref.~\cite{Ruokola11} becomes
wrong, and needs consideration of strong renormalization by the Kondo effect.
Actually, since the energy scale is renormalized from $\hbar \Delta$ to $k_B T_K$,
one should take $\chi_m \sim 2/(k_B T_K)$ to obtain the correct result
of thermal conductance at low temperatures ($T\ll T_K$). 

Finally, we point out that the same cotunneling formula can be derived 
even when the local system is a harmonic oscillator by replacing $\Delta$ with
the frequency of the local oscillator. This indicates that the low-temperature
heat transport of the present model is governed by the fixed-point Hamiltonian
of a local ``harmonic oscillator'' after nontrivial renormalization by the Kondo physics.

\end{document}